\documentclass[twocolumn,aps,prl,showpacs]{revtex4}

\usepackage{graphicx}
\usepackage{times}

\input{MyMc}

\begin{document} 

\title{Semifluxon molecule under control} 

\author{Andreas Dewes}
\email{dewes@uni-tuebingen.de}
\author{Tobias Gaber}
\author{Dieter Koelle}
\author{Reinhold Kleiner}
\author{Edward Goldobin}
\email{gold@uni-tuebingen.de}
\affiliation{
  Physikalisches Institut \& Center for Collective Quantum Phenomena,   Universit\"at T\"ubingen,
  Auf der Morgenstelle 14,
  D-72076 T\"ubingen, Germany
}

% Include the date command, but leave its argument blank.
\date{\today}

\begin{abstract} 

  Josephson junctions with a phase drop $\pi$ in the ground state allow to create vortices of supercurrent carrying only \emph{half} of the magnetic flux quantum $\Phi_0\approx2.07\times10^{-15}\,\mathrm{Wb}$. Such \emph{semifluxons} have two-fold degenerate ground states denoted \state{s} (with flux $+\Phi_0/2$ and supercurrent circulating clockwise) and \state{a} (with flux $-\Phi_0/2$ and supercurrent circulating counterclockwise).
  We investigate a molecule consisting of two coupled semifluxons in a 0-$\pi$-0 long Josephson junction. The fluxes (polarities) of semifluxons are measured by two on-chip SQUIDs. By varying the dc bias current applied to the 0-$\pi$-0 junction, we demonstrate controllable manipulation and switching between two states, \state{sa} and \state{as}, of a semifluxon molecule. These results provide a major step towards employing semifluxons as bits or qubits for classical and quantum  digital electronics.
  
\end{abstract} 


\maketitle 

%Structure: 
%Long Josephson junctions, 0-pi, 0-pi-0 Josephson junctions.
%Magnetic field of fractional vortices, coupling of the vortices to a DC-SQUID.
%Ic(Inj) dependence & simulations.
%Rearrangement processes. Bias current induced rearrangement.
%Discussion of the measurements. Results & simulations.
%Applications, conclusions & outlook.

%Not more than 4 pages and 4 figures.

%List of figures:

%1.) potential energy and phase configuration in 0-kappa-0 JJ 
%2.) microscope image of the samples
%3.) I_c(I_inj)
%4.) a: V_{SQ}(I_{jj}) characteristic. b: V_JJ(I_JJ) characteristic. c: Calculated value of the magnetic flux in the SQUID loop.

%\begin{multicols}{2}%{

\paragraph{Introduction.}

Any introductory book on superconductivity claims that magnetic flux in a superconducting loop is quantized with the single flux quantum being $\Phi_0\approx2.07\times10^{-15}\,\mathrm{Wb}$. The flux quantization owes to the $2\pi$ periodicity of the phase of the wave function of the superconducting condensate and leads to the existence of fluxons --- Abrikosov or Josephson vortices of supercurrent carrying $\pm\Phi_0$. In this case the phase grows smoothly by $\pm2\pi$ if one goes around a loop containing the vortex. It turns out that using a so-called $\pi$ Josephson junction (JJ), i.e. a JJ having a phase drop of $\pi$ in the ground state, embedded in this loop, a phase drop $\pi$ can be ``stolen'' (squeezed inside the $\pi$ junction) and the current circulating in a loop should provide only another $\pi$ or $-\pi$ to have a phase change equal to a multiple of $2\pi$ when one goes around the vortex. As a result, the magnetic flux threading the loop is equal to $\pm\Phi_0/2$ --- a semifluxon or antisemifluxon, respectively. The investigation of semifluxons intensified after the advent of 0-$\pi$ JJs that can be fabricated using d-wave superconductors\cite{Kirtley:SF:T-dep,Hilgenkamp:zigzag:SF,Ortlepp:2006:RSFQ-0-pi}, using conventional superconductors with ferromagnetic barriers\cite{Weides:2006:SIFS-0-pi,DellaRocca:2005:0-pi-SFS:SF} or even creating an artificial discontinuity of the Josephson phase using a pair of tiny current injectors attached to one of the electrodes of a conventional JJ with insulating barrier\cite{Goldobin:Art-0-pi,Buckenmaier:2007:ExpEigenFreq}.

%Long Josephson junctions, 0-pi, 0-pi-0 Josephson junctions.

A 0-$\pi$ JJ consists of two connected parts: a 0 part and a $\pi$ part. The 0 part, if taken separately, would have a phase drop $\mu=0$ in the ground state, while the $\pi$ part would have $\mu=\pi$. If 0 and $\pi$ parts are connected, the ground state phase should ``decide'' which value to take. If both 0 and $\pi$ parts are long enough in terms of the Josephson penetration depth $\lambda_J$ (typically $\lambda_J\sim1\ldots200\,\mathrm{\mu m}$), the ground state phase $\mu(x)$ smoothly changes from 0 to $\pm\pi$ in the $\lambda_J$-vicinity of a 0-$\pi$ boundary\cite{Bulaevskii:0-pi-LJJ,Xu:SF-shape,Goldobin:SF-Shape}. This results in the appearance of a magnetic field $\propto d\mu/dx$ localized in the vicinity of the 0-$\pi$ boundary and a supercurrent $\sim\sin\mu(x)$ circulating around the 0-$\pi$ boundary. The total magnetic flux associated with the localized field (created by a Josephson vortex of supercurrent) is equal to $\Phi=\pm\Phi_0/2$ (if the JJ length $L\gg\lambda_J$). The $\pm$ sign stands for a doubly degenerate ground state with the supercurrent circulating clockwise or counterclockwise\cite{Goldobin:SF-Shape}. Since semifluxons are pinned at the 0-$\pi$ boundary and may have two opposite polarities, which we denote as \state{s} and \state{a}, they can be used in digital electronics to store and process information as classical or quantum bits. 
The physics of single semifluxons has been addressed by several groups theoretically\cite{Bulaevskii:0-pi-LJJ,Xu:SF-shape,Goldobin:SF-Shape,Zenchuk:2003:AnalXover,Susanto:SF-gamma_c,Goldobin:F-SF,Kirtley:IcH-PiLJJ,Lazarides:Ic(H):SF-Gen,Stefanakis:ZFS/2,Nappi:2007:0-pi:Fiske,Goldobin:2KappaEigenModes,Susanto:2005:1D-FractVortexCrystal} and experimentally\cite{DellaRocca:2005:0-pi-SFS:SF,Weides:2006:SIFS-0-pi,Buckenmaier:2007:ExpEigenFreq}. However, in the systems of several \emph{coupled} semifluxons, i.e. \emph{semifluxon molecules}, only the ground states were imaged\cite{Hilgenkamp:zigzag:SF,Kirtley:2005:AFM-SF}. Particularly interesting is the simplest system of two coupled semifluxons in a 0-$\pi$-0 JJ, which has two degenerate ground states \state{sa} and \state{as}. This arrangement was proposed as a candidate for semifluxon based bits and qubits\cite{Kato:1997:QuTunnel0pi0JJ,Goldobin:2005:MQC-2SFs}. It was predicted\cite{Goldobin:SF-ReArrange} that applying a small uniform bias current of positive or negative polarity, one can switch between the \state{sa} and \state{as} states, provided the distance $a$ between semifluxons is of the order of few $(2\ldots5)\lambda_J$. In this report, we demonstrate such a manipulation controlled by a uniform dc bias current and readout of two-semifluxon molecule states using on-chip SQUIDs.

\paragraph{Experiment.}

\begin{figure}[!tb]
  \includegraphics{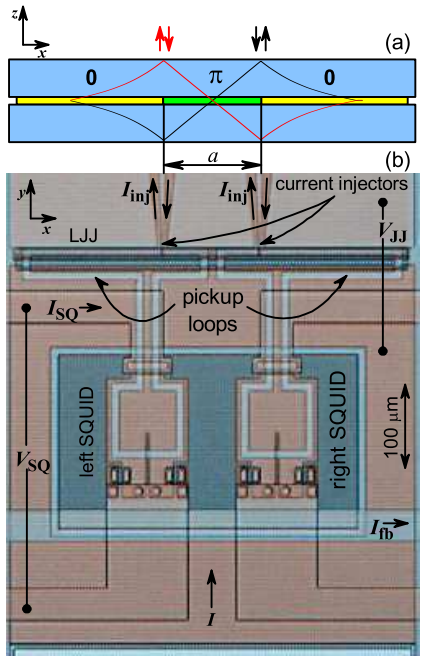}
  \caption{%
    (a) Sketch of the experimental circuit --- a 0-$\pi$-0 LJJ, posessing two degenerate ground states \protect\state{sa} (red) and \protect\state{as} (black). The red and black curves were calculated numerically and represent the local magnetic field $d\mu(x)/dx$. The magnetic field located at each phase discontinuity is coupled to and measured by a dc SQUID.
    (b) Optical microscope image of one of the investigated samples. Two current injector pairs are attached symmetrically to the top superconducting electrode at a distance $a$ (different from sample to sample). The widths $\Delta w$ of each injector and the distance $\Delta x$ between them is $\Delta w\approx\Delta x\approx2\units{\mu m}$. Two dc SQUIDs are placed in front of the phase discontinuities to read out the value of the flux localized at them. Each SQUID is coupled to the long JJ by using a superconducting flux transformer (pickup loop in series with a single turn input coil). 
  }
  \label{Fig:Sketch}
\end{figure}
%
%- coupling loops -> pickup loops\\
%- put (a) and (b) in the top left corner of each figure\\
%- make superconducting electrodes twice thicker in (a) and bluish (sky color)\\
%- DK says that scale of 100microns does not correspond to a=160 microns claimed in the text. Any comments?\\
%- use another (bluish) bitmap for photo. I have sent it to you some time ago.

Fig.~\ref{Fig:Sketch} shows a sketch and a microscope image of one of the samples. To obtain a 0-$\pi$-0 JJ we create two artificial phase discontinuities by using two pairs of current injectors attached to the top superconducting electrode of the JJ at a distance $a$ from each other, see Fig.~\ref{Fig:Sketch}(b). The injector bias circuitry allows to feed both injector pairs by the current $I_\mathrm{inj}$ from a single current source, so that the phase discontinuities created by the injectors have opposite values, i.e. ($\kappa,-\kappa$), where $\kappa\propto I_\mathrm{inj}$. Thus, one obtains a 0-$\kappa$-0 JJ. As soon as $\kappa>0$, the Josephson junction reacts to the phase discontinuities by creating fractional vortices with topological charges that compensate these discontinuities, i.e. $(-\kappa, \kappa)$. For $0<\kappa<2\pi$, one may have two possible stable antiferromagnetically ordered fractional vortex molecules: a $(-\kappa, +\kappa)$ molecule and a $(2\pi-\kappa, \kappa-2\pi)$ molecule. The first one has lower energy for $\kappa<\pi$, the second one for $\kappa>\pi$. For $\kappa=\pi$, these to states become degenerate ground states $(-\pi,\pi)$ and $(\pi,-\pi)$ of the system and are denoted as \state{sa} and \state{as}, respectively. The magnetic field $\propto d\mu(x)/dx$ in both states is shown in Fig.~\ref{Fig:Sketch}(a). The magnetic flux localized around each discontinuity is coupled to one of the two washer-type dc SQUIDs by means of the pickup loop, see Fig.~\ref{Fig:Sketch}(b). The SQUIDs were designed to have a McCumber-Stewart parameter $\beta_c\approx1.1$ and normalized inductance parameter $\beta_L\approx1.5$. The working point of each SQUID was chosen slightly above its critical current. 

The samples were fabricated at Hypres using a $30\,\mathrm{A/cm^2}$ Nb-AlO$_x$-Nb process. They all had $L=480\units{\mu m}$, but different values of $a$. Futher, the results are presented for the sample, which had a length $L=480\units{\mu m}$, and the distance between injectors $a=160\units{\mu m}$. The critical current measured $T=4.2\units{K}$ was  $I_c=1.56\units{mA}$. 

\begin{figure}[!tb]
  \includegraphics{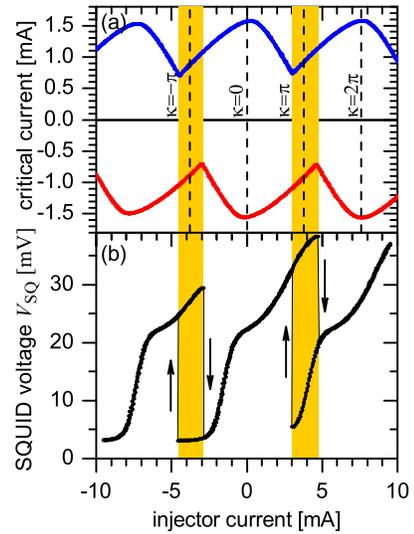}
  \caption{
    %I_c(0)  = 1.56\pm 0.02 mA
    %I_c(pi) = 0.9\pm 0.02 
    (a) $I_{c\pm}(I_\mathrm{inj})$ dependences measured experimentally. The injector current of $I_\mathrm{inj}\approx3.8\,\mathrm{mA}$ corresponds to the point $\kappa=\pi$. 
    (b) The dependence of the SQUID voltage $V_\mathrm{SQ}$ on $I_\mathrm{inj}$ measured experimentally at $I=0$. 
  }
  \label{Fig:Ic(Iinj)}
\end{figure}

First, to calibrate injectors, i.e., to find the value of $I_\mathrm{inj}$ for which $\kappa=\pi$, we have measured the $I_{c}(I_\mathrm{inj})$ dependence, see Fig.~\ref{Fig:Ic(Iinj)}(a). The dependences $I_{c\pm}(\kappa)$ are $2\pi$-periodic in $\kappa$, so that from Fig.~\ref{Fig:Ic(Iinj)}(a) one can immediately say that $I_\mathrm{inj}\approx7.6\,\mathrm{mA}$ corresponds to $\kappa=2\pi$. 
%However, since the $I_{c\pm}(I_\mathrm{inj})$ dependence has rather smooth maxima, such a way to calibrate injectors is not very exact. A more exact procedure of injector calibration is described in Sec.~\ref{Sec:InjCalEx}.
The $I_c(I_\mathrm{inj})$ curves were also calculated numerically for comparison (not shown). The most crucial parameter which is required to compare simulations with the experiment is the normalized JJ length $l=L/\lambda_J$. We have estimated that $l\approx 10.9$ for $\lambda_J$ calculated using conventional expressions, and $l\approx 7.5$ for $\lambda_J$ calculated using idle region corrections\cite{Franz:2000:IdleReg:IcH}. For our JJ geometry the idle region corrections formula should give a bit overestimated values of $l$. Thus, the true value of $l$ should be somewhere in between, but closer to $7.5$. We have found that the simulated $I_c(I_\mathrm{inj})$ fits with the experimental one in the most reasonable way for $L=8.4\lambda_J$ ($a=2.8\lambda_J$), which will be used below. In particular, for such a JJ length, the value of normalized critical current at $\kappa=\pi$ of $0.58$ is in excellent agreement with experiment, where $0.9/1.56\approx 0.58$. 

Second, without biasing the JJ ($I=0$), we have measured the left SQUID voltage $V_L(I_\mathrm{inj})$, see Fig.~\ref{Fig:Ic(Iinj)}(b). The $V_L(I_\mathrm{inj})$ and, therefore, $\Phi_L(I_\mathrm{inj})$, jump at certan values of $I_\mathrm{inj}$, corresponding to abrupt rearrangements of the flux in the molecule. By sweeping the $I_\mathrm{inj}$ in two directions, one can observe regions of bistability around $\kappa=\pm\pi$. In these regions, shown by gold ribbons, both $(-\kappa,+\kappa)$ and $(-\kappa+2\pi,\kappa-2\pi)$ vortex molecules are stable, their energies and fluxes are comparable. The edges of bistable region seem to correspond to the edges of different branches of $I_{c\pm}(I_\mathrm{inj})$ dependence. As we show later, it is just a coincidence. 
%This correspondence is not exact, as starting from one state, e.g. $(-\kappa, \kappa)$, at $I=0$, one may switch to the voltage state from the other state, e.g. $(-\kappa+2\pi, \kappa-2\pi)$, as a result of bias current induced rearrangement, which we address below. 

%
\begin{figure}[!htb]
  \includegraphics{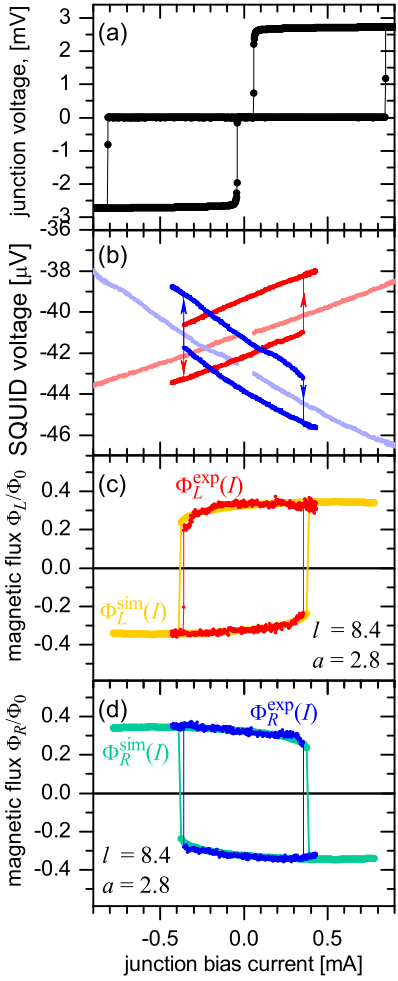}
  \caption{%
    (a) Voltage $V(I)$ across the JJ at $I_\mathrm{inj}=3.8\,\mathrm{mA}$ ($\kappa=\pi$) and (b) corresponding dc SQUID voltages $V_L(I)$ and $V_R(I)$, measured simultaneously. First, the bias current was swept with overcritical amplitude [light red and light blue reference curves in (b), black curve in (a)] and then with undercritical amplitude [red and blue curves in (b), not shown in (a) as $V(I)=0$ in this case]. The small interval of bias currents around $I=0$ for which the JJ was in the Meissner state is cut off from the reference curves in (b). Graphs (c) and (d) show recovered $\Phi_L^\mathrm{exp}(I)$ (red) and $\Phi_R^\mathrm{exp}(I)$ (blue) dependences and their comparison with numerically simulated $\Phi_L^\mathrm{sim}(I)$ (gold) and $\Phi_R^\mathrm{sim}(I)$ (green) curves.
  }
  \label{Fig:Exp}
\end{figure}

Next, we fix $I_\mathrm{inj}=3.8\units{mA}$ ($\kappa=\pi$). A uniform bias current $I$ exerts a Lorenz force acting on the fractional vortices and can induce transitions between the states \state{sa} or \state{as} if the force pushes vortices towards each other\cite{Goldobin:SF-ReArrange}. Such a state transition takes place on the time scale of the inverse plasma frequency $f_p^{-1}$ (in our case $f_p \sim 40\,\mathrm{GHz}$) and will finally result in a reversal of the magnetic flux of both vortices in a molecule.
While ramping the bias current $I$ of the JJ back and forth, we measure the voltage $V(I)$ across the JJ as well as the voltages across each dc SQUID simultaneously, see Fig.~\ref{Fig:Exp}(a,b). First, the bias current $I$ was swept back and forth between the values \emph{exceeding} the critical currents $\pm I_c(I_\mathrm{inj})$, so that most of the time the JJ was in the resistive (McCumber) state, see the black curve in Fig.~\ref{Fig:Exp}(a) and light red and light blue curves in Fig.~\ref{Fig:Exp}(b), further called ``reference curves''. Second, the bias current $I$ was swept back and forth between the values \emph{not exceeding} the positive and negative critical currents $I_{c\pm}(I_\mathrm{inj}^{\pi})$ of the JJ so that the JJ was in the zero voltage state during the whole sweep, see the red and the blue curves in Fig.~\ref{Fig:Exp}(b). One can clearly see that the voltages across both SQUIDs jump simultaneously at $I\approx\pm 0.36\units{mA}$, indicating an abrupt rearrangement of the semifluxon molecule. 

In the resistive state, see the reference curves in Fig.~\ref{Fig:Exp}(b), the net magnetic flux coupled from the molecule to the SQUID averages to zero as a result of fast Josephson oscillations. Nevertheless, the reference curves in Fig.~\ref{Fig:Exp}(b) show an almost linear $V_{L,R}(I)$ dependence, because of the parasitic coupling between the bias current $I$ and the SQUIDs, $\Phi^\mathrm{par}_{L,R}\propto I$. 

The main result of the paper is presented in Fig.~\ref{Fig:Exp}(c,d). It shows the magnetic fluxes $\Phi_{L,R}^\mathrm{exp}(I)$ sensed by the SQUIDs as a function of the (always undercritical) bias current $I$. The fluxes were calculated by subtracting the parasitic flux induced by $I$. To provide better fit, the flux axis was rescaled, which corresponds to the different coupling factors of SQUIDs. In these diagrams, two stable flux states with opposite fluxes $\Phi_{L}^\mathrm{exp}(I)$ and $\Phi_{R}^\mathrm{exp}(I)$ are visible, corresponding to the states \state{sa} and \state{as} of the semifluxon vortex molecule. Transitions between these states were induced at the values of the applied bias current of $I_\mathrm{RE}^\mathrm{exp}\approx0.36\,\mathrm{mA}$. Note that the Josephson junction itself remains in the zero-voltage state during this measurement. The simulated $\Phi_{L,R}^\mathrm{sim}(I)$ dependences are also shown in Figs.~\ref{Fig:Exp}c,d. The simulation was performed in normalized units for experimental parameters ($L$ and $a$) using \textsc{StkJJ}.\cite{StkJJ} The current axis was then rescaled taking into account the measured $I_c$ of the sample. The rearrangement current of $I_\mathrm{RE}^\mathrm{sim}\approx0.38\,\mathrm{mA}$ (in normalized units $I_\mathrm{RE}^\mathrm{sim}/I_c=0.25$) obtained from simulation is in accord with $I_\mathrm{RE}^\mathrm{exp}$, see Figs.~\ref{Fig:Exp}c,d.

%kappa dependence

\begin{figure}[!htb]
  \includegraphics{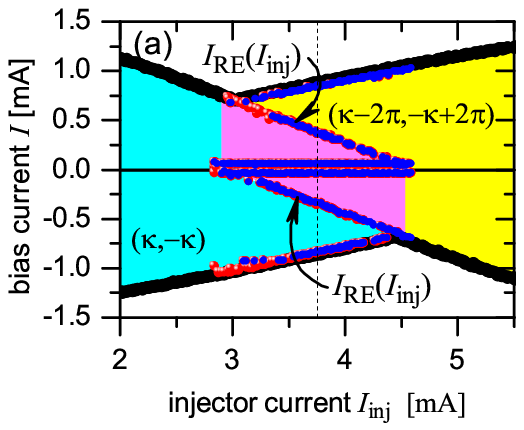}
  \includegraphics{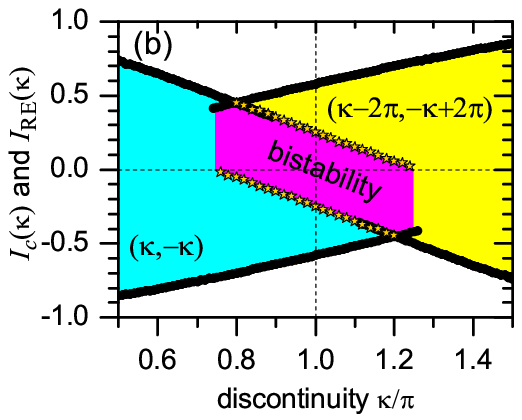}
  \caption{%
    (a) The $I_c(I_\mathrm{inj})$ dependence measured using a voltage criterion algorithm in the vicinity of $\kappa=\pi$ (black symbols). The values of the switching current extracted from experimentally measured $V_{L}(I)$ and $V_{R}(I)$ curves taken at different $I_\mathrm{inj}$ are shown by red and blue symbols. 
    (b) The critical current (black symbols) and rearrangement currents (stars) calculated numerically for $L=8.4\lambda_J$ and $a=2.8\lambda_J$. The region of coexistence of both states (bistability) $(\kappa,-\kappa)$ and $(\kappa-2\pi,-\kappa+2\pi)$ around $\kappa=\pi$ is shown by magenta color.
  }
  \label{Fig:I_RE(I_inj)}
\end{figure}

In our experiment with electronically tunable discontinuity $\kappa\propto I_\mathrm{inj}$ we have a unique possibility to study rearrangements not only in \emph{semifluxon} AFM molecules $(\pi,-\pi)\leftrightarrow(-\pi,\pi)$, but also in molecules made of \emph{arbitrary} fractional vortices $(-\kappa,+\kappa)\leftrightarrow(2\pi-\kappa,+\kappa-2\pi)$, for a range of $\kappa$ in the vicinity of $\kappa=\pi$. To do this, a set of measurements similar to the one outlined above was performed for different values of $\kappa \propto I_\mathrm{inj}$. The values of bias current $I$ at which the SQUID voltages $V_{L,R}(I)$ jump were extracted from $V_{L,R}(I)$ curves using an automatic routine and plotted in Fig.~\ref{Fig:I_RE(I_inj)}(a) as a function of $I_\mathrm{inj}$. The automatic routine was identifying the voltage jumps when a voltage change betwen neighboring points was exceeding a criterion of $0.4\units{\mu V}$, chosen to be just above the voltage noise level.

First, in Fig.~\ref{Fig:I_RE(I_inj)}(a) one can see that both SQUIDs switch mostly simultaneously. In the regions where only the points corresponding to one SQUID are visible the other SQUID is insensitive to flux jumps and produces voltage jumps below the noise level. 
Second, one can observe several branches of the switching current vs. $I_\mathrm{inj}$. One branch corresponds to the switching at $I_c$ and coinsides with the $I_c(I_\mathrm{inj})$ curve also shown in Fig.~\ref{Fig:I_RE(I_inj)}(a). Another two branches, parallel to the $I_\mathrm{inj}$ axis, correspond to switching from the McCumber state to the superconducting state at return current $I=\pm I_R$. Finally, the two most interesting branches marked as $I_\mathrm{RE}(I_\mathrm{inj})$ look like a continuation of the $I_c(I_\mathrm{inj})$ curves towards $I_c=0$. They correspond to a rearrangement between the $(\kappa,-\kappa)$ and $(2\pi-\kappa,+\kappa-2\pi)$ states. The magenta colored region between the $I_\mathrm{RE}(I_\mathrm{inj})$ lines represent a domain where the $(\kappa,-\kappa)$ and $(2\pi-\kappa,+\kappa-2\pi)$ states are both stable. It is this region, which can be used to build various switches, memory devices, bits and qubits. Fig.~\ref{Fig:I_RE(I_inj)}(b) shows numerically calculated $I_c(\kappa)$ and $I_\mathrm{RE}(\kappa)$ dependences that show a state diagram very similar to the experimental one. 

\paragraph{Conclusions.}

We demonstrated experimentally that (a) a molecule of two interacting fractional Josephson vortices with two degenerate states \state{sa} and \state{as} can be created; (b) they can be conveniently manipulated between the \state{sa} and the \state{as} states by applying a small positive or negative dc bias current; and (c) the state of the molecule can be read out using on-chip SQUIDs. These results open a route to the application of fractional flux quanta as convenient information storage devices (superconducting memory). Simultaneously this demonstrates a ready to use preparation and readout technique for a semifluxon molecule based qubit\cite{Kato:1997:QuTunnel0pi0JJ,Goldobin:2005:MQC-2SFs}. 

%Outlook
%Further measurements on similar structures with different arrangement of discontinuities are currently in progress and already provided further evidence on the interaction of the individual fractional vortices in the system. 

Experimentally, one can easily fabricate molecules made of 3, 4, etc. fractional vortices or even a 1D chain (crystal) made of many vortices. The state of these systems can also be controlled by applying a small bias current as presented here. Furthermore, the oscillatory eigenfrequencies of the vortices in the molecules or crystals can be controlled by the same bias current which in this case should be varied within the stability region of each state. Thus, a 1D fractional vortex crystal can be used as a tunable band gap material\cite{Susanto:2005:1D-FractVortexCrystal}, which opens the possibility to realize \emph{tunable} filters in the sub-THz frequency range.

\acknowledgments

Financial support of the DFG (project SFB/TRR-21) is gratefully acknowledged.

\bibliographystyle{apsprl}
\bibliography{this,LJJ,pi,SF,SFS,software}

\end{document}